\documentclass[10pt,twocolumn,floatfix,prl,superscriptaddress]{revtex4-2}
\usepackage{amsmath,amssymb,amsthm,mathrsfs,amsfonts,dsfont,amstext} 
\usepackage{textcomp,pbox}
\usepackage[export]{adjustbox}
\usepackage{bm}
\usepackage{dcolumn,booktabs,url}
\usepackage[scaled]{helvet}
\usepackage{sansmath,gensymb}
\usepackage{tikz,graphicx,transparent,color}
\usepackage{multirow}
\usepackage[separate-uncertainty = true]{siunitx}
\usepackage{comment}
\usepackage{physics}
\usepackage{pdfpages}
\usepackage{array}
\usepackage{makecell}
\usepackage{tabularx}
\usepackage{dsfont}

\usepackage{xcolor}
\newcolumntype{L}{>{\raggedright\arraybackslash}X}
\makeatletter
\AtBeginDocument{\let\LS@rot\@undefined}
\makeatother

\newcolumntype{C}[1]{>{\centering\let\newline\\\arraybackslash\hspace{0pt}}m{#1}}

\usepackage[colorlinks=true]{hyperref}
\usepackage{graphicx}

\hypersetup{
     colorlinks   = true,
     citecolor    = red,
     linkcolor    = blue,
     urlcolor     = red     
}

\graphicspath{{./figs/}}

\usepackage{graphicx}% http://ctan.org/pkg/graphicx

\usepackage{physics}

\begin{document}

\newcommand{\TitleName}{{Independent operation of two waveguide-integrated quantum emitters}}
\title{\TitleName}

\newcommand{\AffCPH}{Center for Hybrid Quantum Networks (Hy-Q), The Niels Bohr Institute, University~of~Copenhagen,  DK-2100  Copenhagen~{\O}, Denmark}
\newcommand{\AffBasel}{Department of Physics, University of Basel, Klingelbergstra\ss e 82, CH-4056 Basel, Switzerland}
\newcommand{\AffUSA}{Present address: Department of
Physics and Astronomy, University of Iowa, 205 N Madison Street, Iowa City, IA 52242, USA}
\newcommand{\AffBochum}{Lehrstuhl f\"ur Angewandte Fest\"orperphysik, Ruhr-Universit\"at Bochum, Universit\"atsstra\ss e 150, 44801 Bochum, Germany}

\author{C. Papon}
\affiliation{\AffCPH{}}
\author{Y. Wang}
\affiliation{\AffCPH{}}
\author{R. Uppu}
\affiliation{\AffCPH{}}
\affiliation{\AffUSA{}}
\author{S. Scholz}
\affiliation{\AffBochum{}}
\author{A. D. Wieck}
\affiliation{\AffBochum{}}
\author{A. Ludwig}
\affiliation{\AffBochum{}}
\author{P. Lodahl}
\affiliation{\AffCPH{}}
\author{L. Midolo}
\affiliation{\AffCPH{}}

\email[Email to: ]{camille.papon@nbi.ku.dk}

\date{\today}

\begin{abstract}
 We demonstrate the resonant excitation of two quantum dots in a photonic integrated circuit for on-chip single-photon generation in multiple spatial modes. The two quantum dots are electrically tuned to the same emission wavelength using a pair of isolated $p$-$i$-$n$ junctions and excited by a resonant pump laser via dual-mode waveguides. We demonstrate two-photon quantum interference visibility of $(79\pm2)\%$ under continuous-wave excitation of narrow-linewidth quantum dots. Our work solves an outstanding challenge in quantum photonics by realizing the key enabling functionality of how to scale-up deterministic single-photon sources.

\end{abstract}

\maketitle 
The coherent generation and manipulation of single photons using solid-state quantum emitters are key resources for the development of scalable quantum information protocols \cite{Uppu2021} to realize quantum simulators \cite{Peruzzo2014,Sparrow2018} or quantum communication hardware \cite{Buterakos,Eva,PhysRevX.10.021071}. Significant progress has been made with InAs self-assembled quantum dots (QDs) embedded in GaAs photonic nanostructures \cite{review2022}, where ultra-high quality epitaxial growth \cite{Bart2022} combined with reproducible low-electrical-noise photonic devices enabless highly coherent photon emission \cite{PhysRevB.96.165306,uppuscience,Tomm2021} and near-transform-limited spectral linewidths \cite{Pedersen2020,senellart_reproducibility}. Embedding QDs in thin membranes further enables direct integration of single-photon sources (SPS) in waveguides, resulting in near-unity light-matter interaction and deterministic source operation, with streams of hundreds of indistinguishable photons being generated \cite{uppuscience}.
  
\begin{figure*}
	\includegraphics{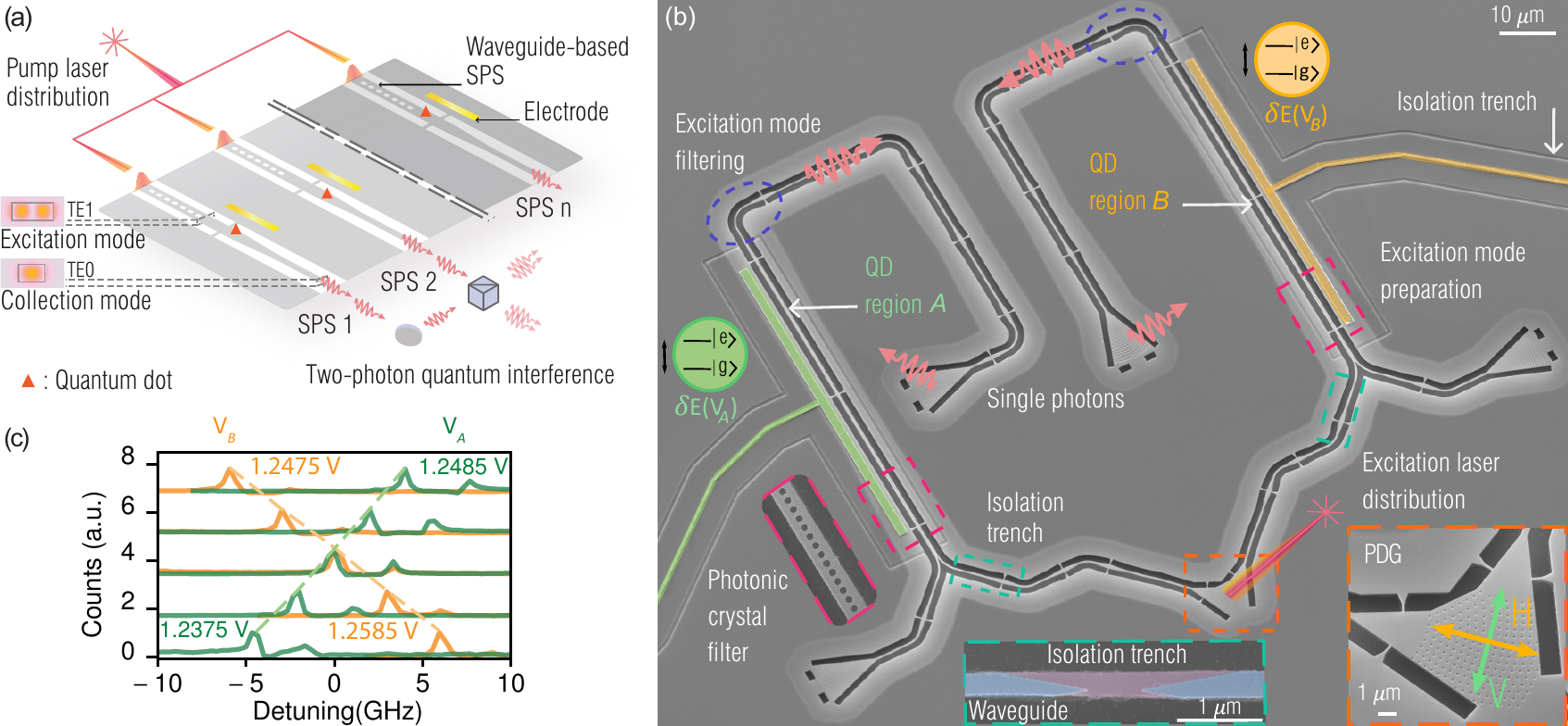} 
	\caption{(a) Sketch of a fully waveguide-integrated scheme for generating single photons in mutiple spatial modes, where a single laser source is distributed on-chip (not shown) to $n$-waveguide-based single-photon sources (SPS). Resonant excitation is performed by a dual-mode waveguide geometry where excitation occurs through the TE1 mode and pump is removed by tapering the waveguide after the quantum dot (QD). Collection of single photons is done via the TE0 waveguide mode. Independent electrical tuning of the QDs accounts for spectral inhomogeneities. (b) Scanning electron micrograph of the actual device designed for $n=2$. A monochromatic laser is coupled through a polarization diversity grating (PDG) (orange dashed area and inset) and is prepared in the mode for resonant excitation by the dual-mode waveguide and the photonic crystal filter (red dashed area and inset). QDs are simultaneously excited in region $A$ and $B$ and are independently biased using two electrodes (green and orange shadow, respectively), isolated by shallow-etched trenches (blue dashed area and inset, purple shadow) on the waveguides (blue shadow) to isolate the $p$-layer of the two QD regions. (c) Resonance fluorescence of two QDs, QD$_A$ and QD$_B$, obtained by sweeping $V_A$ (green curve) and $V_B$ (orange curve), excited with a resonant laser locked at $320.705$ THz. Counts are normalized by the peak value of each QD and shifted for clarity.}
	\label{fig:fig2}
\end{figure*}

{The scalability of the QD platform has previously been demonstrated by interfering photons from remote emitters in distant cryostats \cite{lu300km,Liang2QD,Weber2019}. However, the scalable control and operation of waveguide-coupled quantum emitters within the same photonic integrated circuit have not been reported yet (see \cite{note:SM} for a comparison with Refs. \cite{Ellis,Reindl,Patel2010,Weber2019,Liang2QD,lu300km}).} A major roadblock has been the lack of control over the spectral and spatial distribution of self-assembled QDs \cite{Michler2020}, {together with the absence of a strategy for addressing several QDs simultaneously.} {For these reasons, most multi-photon experiments have relied on spatio-temporal demultiplexing of a single quantum emitter} \cite{lenzini,hummel,20photons}, for which full on-chip integration is highly challenging. Alternatively, the DC Stark effect \cite{WarburtonSTARK} may be employed as the tuning mechanism mitigating the effect of inhomogeneous broadening for the integration of multiple QDs \cite{Petruzzella} in the same photonic circuit. Recently, a SPS was demonstrated using a dual-mode waveguide, whose operational principle exploits one mode for excitation of QDs and the second mode for collection of single photons \cite{Uppu2020}.{With this method a single resonant laser may be distributed on-chip to excite multiple  waveguide-integrated quantum emitters, see Fig. \ref{fig:fig2}(a),  thus offering a route to large-scale integration, ultimately enabling on-chip quantum information protocols.}

{In the present work, we demonstrate the independent operation of two quantum emitters using a tailored nanophotonic circuit, whose design principle can be further scaled to a larger number of emitters.} We realize simultaneous resonant excitation of two QDs positioned in separate dual-mode waveguides and demonstrate their independent Stark tuning with electrical biases applied locally. The device is based on a $p$-$i$-$n$ GaAs membrane containing QDs in its center and individual frequency tuning is implemented by fabricating local electrical contacts. {A polarization diversity grating (PDG) \cite{VanLaere:09} allows distributing the same excitation laser to the individual waveguides and equalizing the Rabi frequency of each QD in units of their respective decay rate. The} single-photon character of the emission is analyzed from second-order correlation measurements of each QD under continuous-wave (cw) resonant excitation. Finally, two-photon quantum interference (TPQI) between two independently-controlled QDs is realized, which is a key result demonstrating the scalability of the platform to multiple emitters.

A scanning electron microscope image of the device is presented in Fig. \ref{fig:fig2}(b). The footprint is only $100\times 125 \  \mu\text{m}^2$ providing a compact building block, adoptable to larger number of resonant QDs. The entire structure has been fabricated on gated GaAs membranes (see Supp. Mat. of Ref.~\cite{Uppu2020} for details) with a QD density of $10\ \mu$m$^{-2}$, ensuring a high probability of finding two QDs Stark-tunable into mutual resonance \cite{note:SM}. {A monochromatic laser is coupled to the device via a PDG (orange dashed area in Fig.~\ref{fig:fig2}(b)) distributing optical power into two entrance waveguides, with a ratio controlled with the incident light's polarization. The PDG, inspired from \cite{VanLaere:09}, has been entirely redesigned and optimized to operate at $930$ nm on suspended GaAs waveguides \cite{note:SM}.} The laser is initially prepared in a superposition of the fundamental (even symmetry) and first-order (odd symmetry) modes of a dual-mode waveguide by an inverted Y-splitter and then filtered by a one-dimensional photonic crystal mirror (inset of the red dashed area in Fig.~\ref{fig:fig2}(b)) so that only the odd mode reaches the QD regions \cite{Uppu2020}. The single photons are collected into the fundamental TE0 mode of the waveguide and routed towards the out-coupling shallow-etched grating (SEG) \cite{xyz}, while the photonic crystal sections back-reflect any photons emitted towards the excitation port. The excitation laser is filtered by the first 90-degree bend (dashed purple circle in the figure), where the waveguide is adiabatically tapered. Independent Stark tuning of QDs in each waveguide is achieved by applying a homogeneous vertical electric field using electrodes (highlighted in orange and green) contacted to the $p$-layer of the heterostructure. A global $n$-back contact is fabricated in the periphery of the structures. {A set of $50$-nm-deep isolation trenches surrounding the electrodes and crossing the waveguides (see blue dashed areas in Fig.~\ref{fig:fig2}(b)), previously reported for reducing the $p$-$i$-$n$ diode capacitance and response time \cite{Pedersen2020,martin_prl}, have been adapted to the composite dual-mode waveguides device, to achieve isolated tuning of the two QD regions. The tapered geometry of the trenches ensures minimal optical losses}.  

The device is characterized at cryogenic temperature (T=$1.6$ K) by coupling a cw laser through the PDG and sweeping the bias voltage over the two QD regions. The resonance fluorescence (RF) collected from QD region $A$ and $B$ is detected with two superconducting nanowire single-photon detectors (SNSPDs). Figure~\ref{fig:fig2}(c) shows the RF signal as a function of the frequency detuning between two different QDs defined as QD$_A$ (green curve) and QD$_B$ (orange curve) for a laser frequency locked at $320.705$ THz. From the full frequency-voltage plateau lines \cite{note:SM} in the Coulomb blockade regime \cite{Warburton2013,single_charge}, we attribute these states to the QD neutral exciton. The two transitions are brought in mutual resonance by sweeping simultaneously $V_A$ and $V_B$, the voltage on waveguide $A$ and $B$, respectively, without crosstalk \cite{note:SM}. Both QDs have a weak second transition frequency-shifted by approximately 3 GHz, attributed to the second dipole of the neutral exciton \cite{uppuscience}, whose coupling efficiency to the waveguide depends on the QD lateral position \cite{note:SM,Rotenberg:17,fdtd,meep}. 

\begin{figure}
	\includegraphics{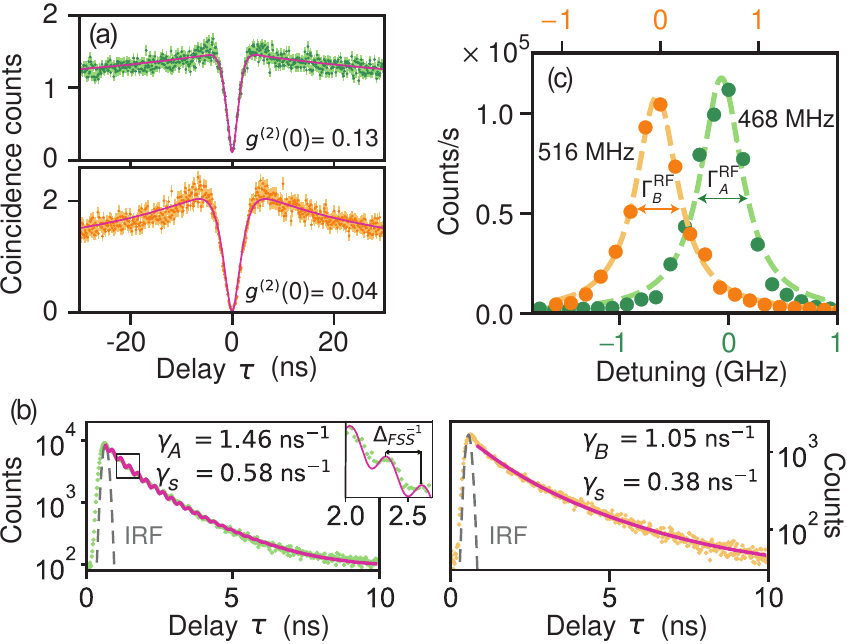} 
	\caption{ (a) Second-order correlation functions of QD$_A$ (top) and QD$_B$ (bottom). The pink curves show the fit, from which we extract the radiative decay of each QD. (b) Lifetime of QD$_A$ (left) and QD$_B$ (right) are recorded under p-shell excitation. The slow (fitted) and fast (fixed) decay rate of a double-exponential (pink curves) is shown for both QDs. IRF: instrument response function. The fit to QD$_A$ includes an oscillating term from the second dipole, with $\Delta_{FSS}/2\pi = 3.45 \pm 0.01$ GHz. (c) {RF scattered intensity as a function of laser detuning} for QD$_A$ (green dots) and QD$_B$ (orange dots), below saturation. The dashed lines are the fit to a Lorentzian. }
	\label{fig:fig3}
\end{figure}

The single-photon nature of the collected signal is confirmed by the measurement of the second-order correlation function, $g^{(2)}(\tau)$, shown in Fig.~\ref{fig:fig3}(a), at the voltages $V_A$ and $V_B$ ensuring mutual resonance between the two QDs. The time correlation between two SNSPDs is recorded for QD$_A$ and QD$_B$, and reveals a value at zero-time delay of  $g^{(2)}_A(0)=0.13 \pm 0.02$ (top) and $g^{(2)}_B(0) = 0.04 \pm 0.02$  (bottom), respectively, limited by the finite laser suppression \cite{note:SM,impurity}{ sensitive to the fabrication reproducibility  of the structure.} Coincidence counts are normalized by the average value at long time delay and errorbars are estimated from Poissonian statistics. The $g^{(2)}(\tau)$ functions are fitted (pink curves in Fig.~\ref{fig:fig3}(a)) with a model for resonant $g^{(2)}(\tau)$ \cite{Flagg2009}, which includes an exponential term accounting for the bunching observed at short time delay \cite{BlinkingMichler} and the convolution with the instrument response function of the SNSPDs. Radiative decay rates of $\gamma_{A} = (1.46\pm0.03)$ ns$^{-1}$ and  $\gamma_{B} = (1.05\pm0.01)$ ns$^{-1}$ are extracted from the fit, for QD$_A$ and QD$_B$ respectively, after fixing the Rabi frequency to $\Omega_A = 0.48 \gamma_{A}$ and $\Omega_B = 0.34\gamma_{B}$ from power saturation measurement  \cite{note:SM,Muller}. {By controlling the polarization of the incident laser on the PDG, we thus ensure similar driving conditions for each QD.} {We attribute the bunching at short time delay  to the effect of spectral diffusion, visible in  cross-correlation measurements with a narrow-bandwidth laser \cite{Sallen2010}}.

To further characterize the two QDs, time-resolved fluorescence under p-shell excitation through the waveguide is recorded, after spectral filtering of the fluorescence. The decay curves are shown in Fig.~\ref{fig:fig3}(b) and are fitted with a double-exponential, where the fast decay rate is fixed to $\gamma_{A}$ (QD$_A$, left) and $\gamma_{B}$ (QD$_B$, right). For QD$_A$, the presence of the second dipole is highlighted by the beating in the decay curve, with a fine structure splitting of $\Delta_{FSS}/2\pi = 3.45 \pm 0.01$ GHz, in agreement with the RF spectra shown in Fig.~\ref{fig:fig2}(c). The fit also reveals a slow decay component attributed to the pronounced second dipole. Similarly, the fit of the decay curve of QD$_B$ also shows a slow component, which can be attributed to interaction with non-radiative states \cite{Johansen}. From the agreement between the measured decay curves and the model, we extract the lifetime-limited contribution to the linewidth of $\Gamma_{A} =\gamma_{A}/2\pi= 233$ MHz and  $\Gamma_{B} =\gamma_{B}/2\pi=  167$ MHz for QD$_A$ and  QD$_B$, respectively. 

A fine-step RF scan of the two QDs excited simultaneously is shown in Fig.~\ref{fig:fig3}(c) and fitted with Lorentzian functions giving similar linewidths of $\Gamma^\text{RF}_A=468 \pm 28$ MHz and $\Gamma^\text{RF}_B=516 \pm 13$ MHz, for QD$_A$ and  QD$_B$ respectively. Broadening beyond the transform limit ($\Gamma_A$, $\Gamma_B$) is attributed partly to power broadening \cite{Flagg2009} as highlighted in \cite{note:SM}. In the low-power limit, the linewidth of QD$_A$ (QD$_B$) is $1.2\Gamma_{A}$ ($2.6\Gamma_{B}$), due to additional contribution from slow spectral diffusion. This broadening is further characterized by fitting the low-power RF data to a Voigt function, thereby extracting the Gaussian distribution of spectral diffusion with standard deviation $\sigma_A = 68 \pm 4$ MHz ($\sigma_B = 163 \pm 25$ MHz) \cite{note:SM}. The effect of slow noise is comparable to previously recorded RF measurements on the same platform \cite{Uppu2020,uppuscience}. The electrically-contacted sources have a two-fold advantage: the individual tuning allows to bring two different QDs on resonance and the control of the charge environment reduces significantly the impact of charge noise \cite{Kuhlmann2015}, as observed from the narrow linewidths. Even narrower optical linewidths have previously been reported for the planar nanophotonic waveguide platform  \cite{Pedersen2020}, which confirms the potential of the approach. 

\begin{figure}[h!]
 	\includegraphics{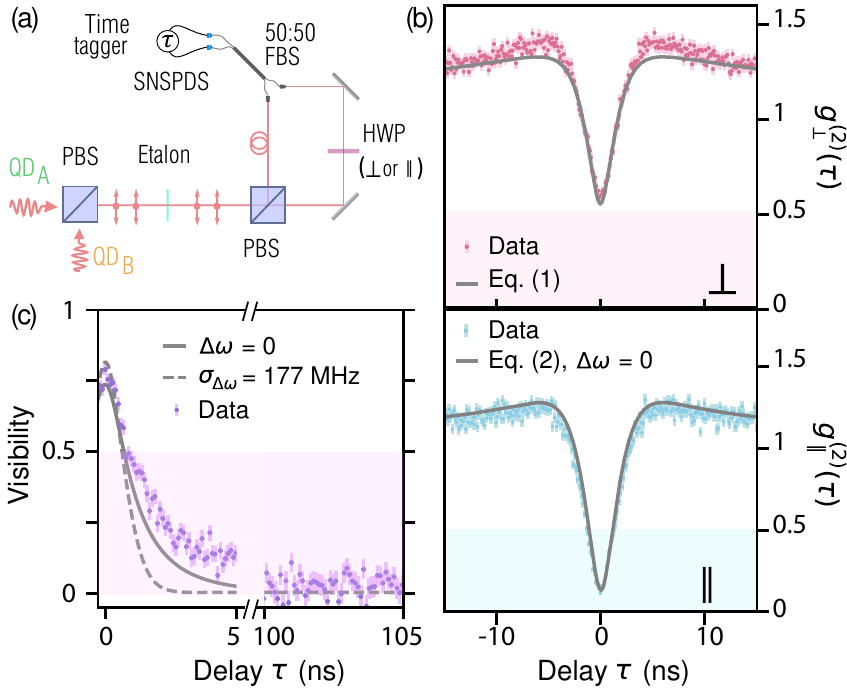} 
 	\caption{(a) Schematic of the interferometer for two-photon quantum interference (TPQI) measurements (PBS: polarized beam splitter, FBS: fiber beam splitter, HWP: half-wave plate, SNSPDs: superconducting nanowire single-photons detectors). (b) Top: Measured $g^{(2)}_{\perp}(\tau)$ for cross-polarized photons (pink dots) with the prediction of Eq.~(\ref{eq:g2perp}) (gray curve). Bottom:  Measured $g^{(2)}_{\parallel}(\tau)$ for co-polarized photons (blue dots) with the prediction of Eq.~(\ref{eq:g2par}) for $\Delta\omega=0$ (gray curve). The shaded regions indicates the threshold below which non-classical effects occur. (c) Visibility of the TPQI (purple dots). The plain (dashed) gray curve is the visibility calculated with $\Delta\omega=0$ (by averaging over a random distribution of width $\sigma_{\Delta\omega} = 177$ MHz). The shaded region indicates the threshold above which non-classical effects occur. }
 	\label{fig:fig4}
 \end{figure}

The low-noise characteristic of both resonant QDs enables testing TPQI. We perform a Hong-Ou-Mandel (HOM) experiment with a balanced Mach-Zender-Intererometer (MZI) under cw-excitation \cite{Proux}, using the same excitation power condition as for the measurement of $g^{(2)}(\tau)$. The HOM setup is shown in Fig.~\ref{fig:fig4}(a). The photons collected from QD$_A$ and QD$_B$, in orthogonally-polarized modes, are combined on a polarized beam splitter and spectrally filtered ($3$ GHz linewidth) to remove phonon sidebands \cite{PhononPetru} before entering the MZI. When the photons are orthogonally-polarized, i.e. fully distinguishable, the probability $g^{(2)}_{\perp}(\tau)$ of detecting photons on both detectors as a function of the time delay $\tau$ between the two events is \cite{Sandoghdar,Patel2010}
\begin{align} 
\label{eq:g2perp}
\begin{split}
g^{(2)}_{\perp}(\tau) &= c_A^2g^{(2)}_A(\tau)+c_B^2g^{(2)}_B(\tau)+2c_Ac_B,
\end{split}
\end{align}
where $c_n=I_n/(I_A+I_B)$, $I_n$ being the the intensity recorded for QD$_n$ (for $n \in \{A,B\}$), characterized over the HOM measurement to be, on average, $c_A=0.59$ and $c_B=0.41$. The cross-polarized measurement is shown in Fig.~\ref{fig:fig4}(b) (pink dots, top) together with the prediction from Eq.~(\ref{eq:g2perp}) (gray curve) calculated with the fitted $g^{(2)}_n(\tau)$ from Fig.~\ref{fig:fig3}(a). 

TPQI is observed when both arms of the MZI have the same polarization, which results in a vanishing coincidences at small $\tau$. This measurement is shown in Fig.~\ref{fig:fig4}(b) (blue dots, bottom) with an observed TPQI dip reduced much below the classical threshold of 0.5. The TPQI can be modelled with \cite{Sandoghdar,Patel2010}
\begin{multline} 
\label{eq:g2par}
g^{(2)}_{\parallel}(\tau)= c_A^2g^{(2)}_A(\tau)+c_B^2g^{(2)}_B(\tau)+2Rc_Ac_B\times\\\Big[1-\zeta_A\zeta_B|g^{(1)}_A(\tau)||g^{(1)}_B(\tau)|\cos{(\Delta\omega\tau)}  \Big],
\end{multline}
where $R$ is a constant to ensure that $\lim_{|\tau|\rightarrow\infty }g^{(2)}_{\parallel}(\tau)=1$ \cite{Woolley_2013}  (see \cite{note:SM} for details). The residual laser photons account for imperfect interference through $\zeta_n=\sqrt{1-g^{(2)}_n(0)}$. The first-order coherence function $g^{(1)}_n(\tau)$ of  QD$_n$ is a function of the lifetime of the emitters, Rabi frequency and dephasing \cite{scully_zubairy_1997} and is calculated for this experiment with the measured parameter $\gamma_A$, $\gamma_B$, $\Omega_A$ and $\Omega_B$ \cite{note:SM}. The dephasing rate is here neglected since it is not measured directly. Moreover, previous work in similar conditions shows that it contributes moderately to the loss of interference visibility \cite{Uppu2020}. For QDs fully resonant over the course of the experiment, we have $\cos{(\Delta\omega\tau)=1}$, where $\Delta\omega$ is the frequency detuning between the two emitters. The plain gray curve in Fig.~\ref{fig:fig4}(b) (bottom) is the expected $g^{(2)}_{\parallel}(\tau)$ for $\Delta\omega=0$, where only the normalization constant $R$ is fitted \cite{note:SM,Gerber_2009}. The measured TPQI is well described by the model, however, the time response is slower than from calculation, indicating an overestimation of the decay rate and/or Rabi frequency. This can be corrected in the future by directly measuring the first-order coherence.
 
The visibility of the HOM measurement, defined as $V(\tau)= 1-g^{(2)}_\parallel(\tau)/g^{(2)}_\perp(\tau)$, is displayed in Fig.~\ref{fig:fig4}(c) and reveals a raw peak of $(79\pm2) \%${, primarily limited by the non-zero $g^{(2)}(0)$ of each QD}. The visibility calculated with  $\Delta\omega=0$ (plain gray curve), shows a significant overlap with the measured data, both in width, governed by the dynamics of the QDs, and in peak value, limited by the laser leakage measured in Fig.~\ref{fig:fig3}({a}). Surprisingly, although low-power RF scans indicate the presence of slow spectral diffusion for both QDs, the TPQI dip and hence the visibility peak are well explained without taking this effect into account. To interpret this result, we calculate the visibility curve assuming that the spectral diffusion experienced by the two QDs is fully uncorrelated, meaning that the detuning $\Delta\omega$ varies stochastically. In this case, the mutual detuning, on average $\overline{\Delta\omega}=0$, would follow a normal distribution with $\sigma_{\Delta\omega} = 177$ MHz, given by $\sqrt{\sigma_{A}^2+\sigma_{B}^2}$ \cite{note:SM}, from which we perform an ensemble average of Eq.~\ref{eq:g2par}. We observe a clear deviation from this model in term of the peak width, as shown in Fig.~\ref{fig:fig4}(c) (dashed gray curve), indicating that the dominant slow noise sources are not fully uncorrelated for independent QDs. This may be a result of the two QDs being biased by close-by electrodes thereby featuring similar electrical-noise properties. We note that the calculated peak value of the visibility curves differs due to the uncertainty added by the ensemble average and the normalization of the TPQI expression. The narrow bandwidth of the laser suppression from the current device prevented the implementation of triggered excitation with large signal-to-noise ratio, and hence the direct measurement of indistinguishability through pulsed HOM experiment. Next-generation devices will focus on improving the performance of the laser filtering, in terms of bandwidth and fabrication imperfections.
 
In this work, we showed the operational principle of a multi-QD photonic circuit for scaling up to multiple {quantum emitters}. We demonstrated the simultaneous resonant excitation of two QDs and their independent wavelength control. Laser suppression was achieved by the waveguide design and was confirmed by the measurement of strong anti-bunching in the second-order correlation function. We have analyzed two QDs in detail, both revealing narrow and similar emission linewidths showing the high coherence of the photon-emitter interface. We studied the presence of slow residual spectral diffusion and found that the effect on TPQI was limited due to correlated noise on the two QDs, i.e. the detrimental effect may be less severe than anticipated. It is also noted that residual slow noise can be further reduced by frequency-locking of the emitters \cite{lock}. Another route to improvement exploits Purcell enhancement{by slow-light and dispersion engineering} in a dual-mode photonic-crystal waveguide \cite{Zhou_2022}, as a direct extension of the present experiment. Furthermore, we showed that two {quantum emitters} can be conveniently controlled simultaneously, which, {combined with pulsed excitation,} represents the necessary hardware {for coherent multi-emitter interaction \cite{lx_Science} and }device-independent quantum key distribution, requiring high source efficiency \cite{Eva}. Moreover, scaling up to more sources could be realized by cascading a network of tunable beam-splitters \cite{Papon:19} for pump distribution, and tunable filters \cite{xiaoyan_filter} could be integrated for phonon sideband suppression. This resonant excitation technique can also be adapted to different quantum emitter platforms, such as GaAs QDs, where high indistinguishability between remote QDs has been recently demonstrated \cite{Liang2QD}. 
 
\begin{acknowledgments}
\subsection*{Acknowledgments}
We thank A. Tiranov for experimental assistance and E. M. Gonz\'alez-Ruiz and A. S. S\o rensen for guiding discussion on two-photon quantum interference. We thank Baptiste Karoubi for assistance in designing the concept figure. We gratefully acknowledge financial support from the Danish National Research Foundation (Center of Excellence Hy-Q DNRF139), Styrelsen for Forskning og Innovation (FI)
(5072-00016B QUANTECH), Innovationsfonden (No. 9090-00031B, FIRE-Q), and the European Research Council (ERC) under the European Union’s Horizon 2020 research and innovation programme (Grant agreement No. 949043, NANOMEQ). A.L., S.S., and A.D.W. acknowledge support of BMBF (QR.X Project 16KISQ009) and the DFG (383065199 and TRR 160/2-Project B04).
\end{acknowledgments}

%\bibliography{reflist}% Produces the bibliography via BibTeX.
%apsrev4-2.bst 2019-01-14 (MD) hand-edited version of apsrev4-1.bst
%Control: key (0)
%Control: author (8) initials jnrlst
%Control: editor formatted (1) identically to author
%Control: production of article title (0) allowed
%Control: page (0) single
%Control: year (1) truncated
%Control: production of eprint (0) enabled
%

\end{document}

\begin{center}
		\textbf{\large Supplemental Materials: {Independent operation of two waveguide-integrated quantum emitters}}
	\end{center}
	%%%%%%%%%% Merge with supplemental materials %%%%%%%%%%
	%%%%%%%%%% Prefix a "S" to all equations, figures, tables and reset the counter %%%%%%%%%%
	\setcounter{equation}{0}
	\setcounter{figure}{0}
	\setcounter{table}{0}
	\setcounter{page}{1}
	\makeatletter
	\renewcommand{\theequation}{S\arabic{equation}}
	\renewcommand{\thefigure}{S\arabic{figure}}

	%%%%%%%%%% Prefix a "S" to all equations, figures, tables and reset the counter %%%%%%%%%%

	\section{Success probability of the resonant excitation of a quantum dot pair}
\begin{figure*}[h!]
		\includegraphics[width=0.5\textwidth]{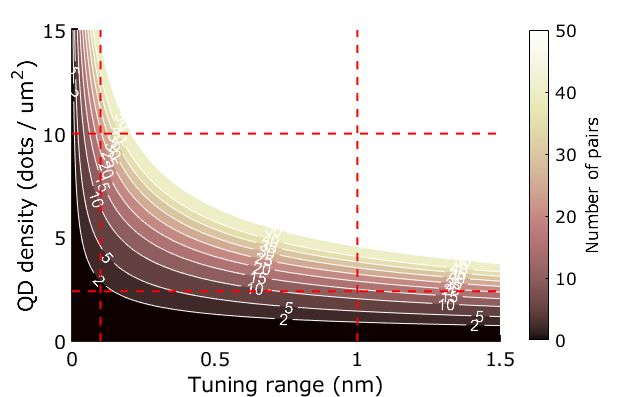}
		\caption{Number of QD pairs as a function of the wavelength tuning range and QD density, in $40 \ \mu$m long and $200$ nm wide waveguide. The wavelength distribution of self-assembled quantum dots is assumed to follow a normal distribution of $15$ nm standard deviation. The QD density is controlled by the growth technique and the tuning range is set by the layout of the heterostructure.}
		\label{fig:density}
	\end{figure*}

	Self-assembled quantum dots are randomly distributed in position and wavelength, meaning that the fabrication of single-photon sources is not deterministic. Instead, we can estimate the probability of finding a pair of resonant QDs in two different structures given the QD density, controlled by the epitaxial growth technique, and the dc Stark shift tuning range, limited by the layout of the wafer heterostructure.  By assuming that the inhomogeneous broadening of the QDs follows a Gaussian wavelength distribution with central wavelength $930$ nm and a standard deviation of $15$ nm, the probability of finding two QDs with wavelengths within the tuning range $\delta\lambda$ is expressed as
	\begin{align} 
		\label{eq:probability}
		P(0\leq\lambda\leq\delta\lambda)=\int_{0}^{\delta\lambda}\frac{1}{\sigma}\sqrt{\frac{2}{\pi}}e^{\frac{-\lambda^2}{2\sigma^2}}.
	\end{align}
	Depending on the area of the device $A$ and the QD density $\rho_\text{QD}$, the total number of QD pairs is $(A\cdot \rho_\text{QD})^2$. Finally, the number of QD pairs within $\delta\lambda$ is calculated as $P(0\leq\lambda\leq\delta\lambda)\cdot(A\cdot \rho_\text{QD})^2$. The result is given in Fig.~\ref{fig:density}, for different tuning range $\delta\lambda$ and QD density. We added a $50\%$ penalty on the number of QD pairs, to account for the cavity effect due to reflection from the filter mirror and finite reflection of the shallow-etched couplers \cite{xyz}. In this work, with a tuning range of $0.1$ nm and a QD density of $10\  \mu \text{m} ^{-2}$, we expect to find approximately $25$ pairs, given a length of the QD region of $40\ \mu$m. In the investigated device we found $3$ pairs. The discrepancy is likely due to limited bandwidth of the photonic crystal filter, linked to fabrication disorder. Larger wavelength tuning range, up to $1$ nm, are achievable with modified wafer heterostructures \cite{PhysRevB.96.165306} and will increase the number of QD pairs. {Finally, the fabrication yield of suspended waveguide-integrated single-photon sources is an important parameter of success probability, and here we report $100 \%$ survival rate of the devices after fabrication.}
	
	\section{Characterization of the polarization diversity grating}
\begin{figure*}[h!]
		\includegraphics{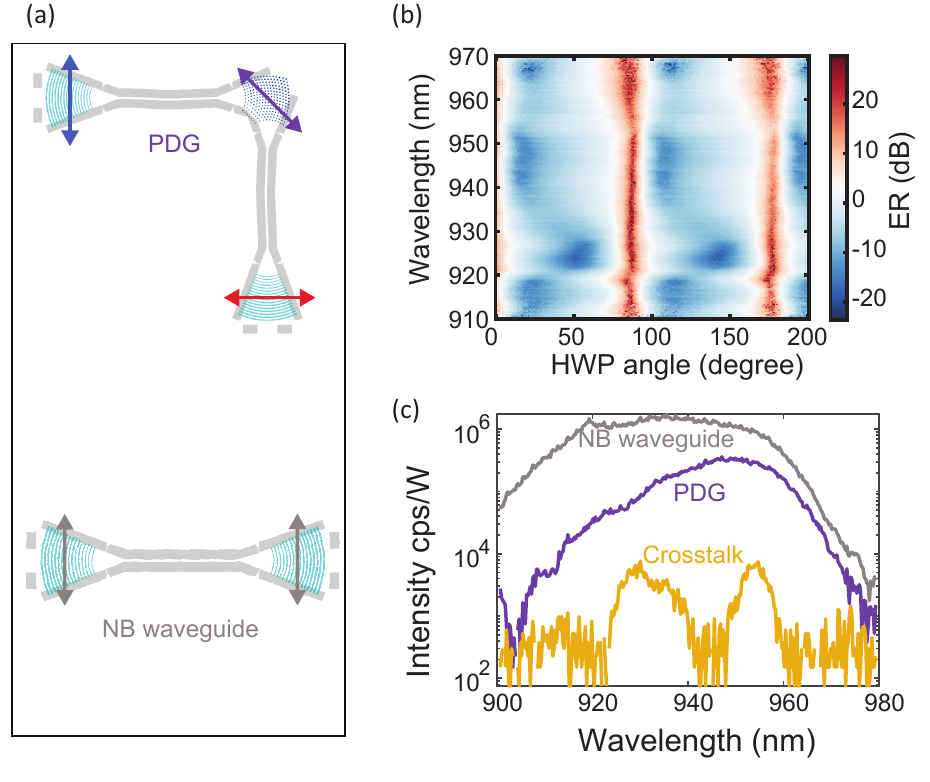}
		\caption{Characterization of the polarization diversity grating. (a) For characterization and alignement, a PDG is coupled to two single-mode waveguides terminated by orthogonally polararized grating couplers. For coupling efficiency reference, a single-mode nanobeam waveguide terminated by two SEGs is included. (b) Exctincion ratio between the two grating couplers as a function of the polarization of the incident light on the PDG for different wavelengths. (c) Transmission through a nanobeam waveguide compared to the transmission through a PDG when the polarization is maximized for the "blue" outcoupler. Crosstalk is measured by coupling in the "red" grating coupler and collecting from the "blue". Measurements are performed at $10$ K.}
		\label{fig:PDG}
	\end{figure*}

	The polarization diversity gratings (PDG) are designed according to Ref. \cite{VanLaere:09}. To this end, we superimpose the ellipses of two perpendicular shallow-etched gratings (SEG) \cite{xyz} and define holes at the intersection. The PDG are then fabricated during the same lithography and shallow-etching step as for the SEG and the shallow trenches. To characterize the transmission of the PDG, a simple test structure is defined, where the waveguides are terminated by SEG, as shown in Fig.~\ref{fig:PDG}(a,top). A super-continuum laser light (SuperK) is coupled to the PDG and the intensity at the left (blue port) and botttom (red port) SEG is recorded as a function of wavelength and polarization of the input. A clear broadband tuning of the coupling to each waveguide is seen in Fig.~\ref{fig:PDG}(b), with extinction ratio larger than $20$ dB in absolute value. The transmission through the PDG is compared to the one through a normal nanobeam waveguide (gray curve), as seen in Fig.~\ref{fig:PDG}(a,bottom), when the coupling on the PDG is maximized for the blue outcoupler (purple curve). By comparing the transmission, we estimate a maximum diffracting efficiency of  $40 \%$ compared with a SEG at $950$ nm, which is limited by mode overlap and etching depth. The crosstalk between the two output ports is negligible, characterized by the transmission from the blue port to the red port, shown as the yellow curve in Fig.~\ref{fig:PDG}. 
	\section{Charge plateaus}
\begin{figure*}[h!]
		\includegraphics{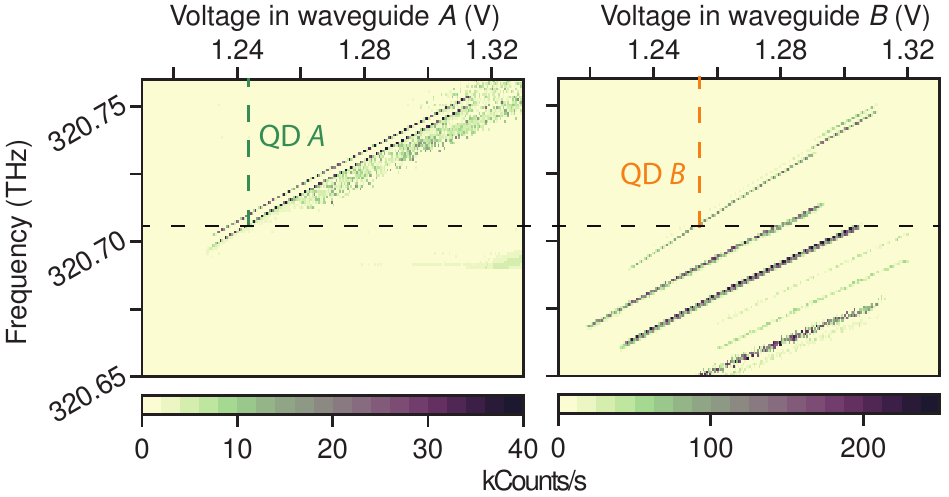}
		\caption{Resonance fluorescence collected from waveguide $A$ (left) and waveguide $B$ (right) as a function of the resonant continuous wave laser frequency and voltage applied on each waveguide. The black dashed line indicates the laser frequency for the experiment described in the main text and the green and orange dashed lines indicate the corresponding resonant voltage for QD$_A$ and QD$_B$, respectively.  }
		\label{fig:freq}
	\end{figure*}

	Measuring the resonance fluorescence (RF) signal as a function of the frequency of a continuous-wave (cw) excitation laser and of the voltage applied to waveguide $A$ and $B$ reveals several plateau lines, shown in Fig.~\ref{fig:freq}.
	\newline
	These plateaus are typical of individual energy-tunable excitonic transitions from QDs in the Coulomb blockade regime \cite{Warburton2013} whose frequency and voltage range are characteristic of the neutral exciton in this heterostructure \cite{uppuscience}. The presence of two lines tuning parallely in waveguide $A$ (left) indicates that the two dipoles of the neutral exciton of QD$_A$ couple to the waveguide mode, as explained in section \ref{seq:beta}. There is a similar second dipole for QD$_B$, but with weaker coupling. The dashed black curve indicates the frequency of the resonant laser for the experiment in the main text while the green and orange lines mark the voltage at which both QDs are mutually resonant. {The charge plateau of QD$_B$ shows a shift in energy at higher voltages, which we attribute to the presence of single charge fluctuation in the local environment of the quantum dot \cite{single_charge}}. We note that several QDs can also be excited resonantly in waveguide $B$, among which two more QDs could be tuned in resonance with QD$_A$. {From previous characterization of quantum dots on the same wafer, under the same experimental conditions, we estimate the emission into the zero-phonon line to be $95\%$ \cite{uppuscience}.}
	
	\section{Electrical isolation from the shallow-etched trenches}
\begin{figure*}[h!]
		\includegraphics{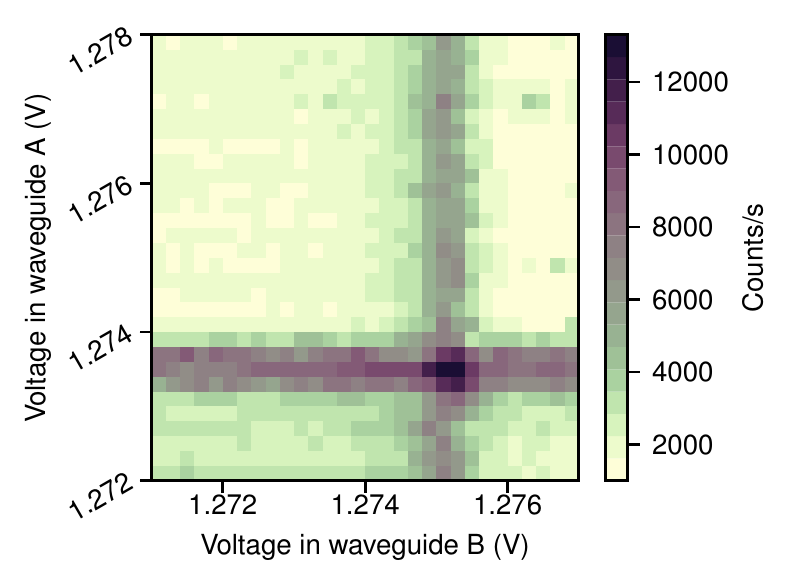}
		\caption{Resonance fluorescence scans of two QDs from a different structure than investigated in the main text, as a function of voltages on both waveguides. We add the RF signals from both QDs to visualize them together.}
		\label{fig:trench}
	\end{figure*}

	Shallow-etched trenches were used in a previous work in order to reduce the RC constant of local $p$-$i$-$n$ diodes \cite{Pedersen2020}, a demonstration of good electrical isolation. In this work, shallow-etched trenches are employed to ensure individual tuning of QDs in different waveguides without crosstalk. In Fig.~\ref{fig:trench}, we show a measurement of RF for two QDs at a fixed frequency of the laser, in a copy of the structure discussed in the main text, as a function of voltages in waveguide $A$ and waveguide $B$. From this two-dimensional RF map we conclude that QDs can be tuned independently without crosstalk, since there is  no frequency shift in QD$_A$ due to the frequency applied in waveguide B (and vice versa). Additionally, we note that the lineshape remains constant for both QDs, which proves that simultaneous tuning can be performed without adding electrical noise. We expect the trenches to work in similar way for the structure investigated in the main text, given the uniformity of the fabrication of the shallow-etched features.
	
	\section{$\beta$-factor calculation}
	\label{seq:beta}
\begin{figure*}[h!]
		\includegraphics{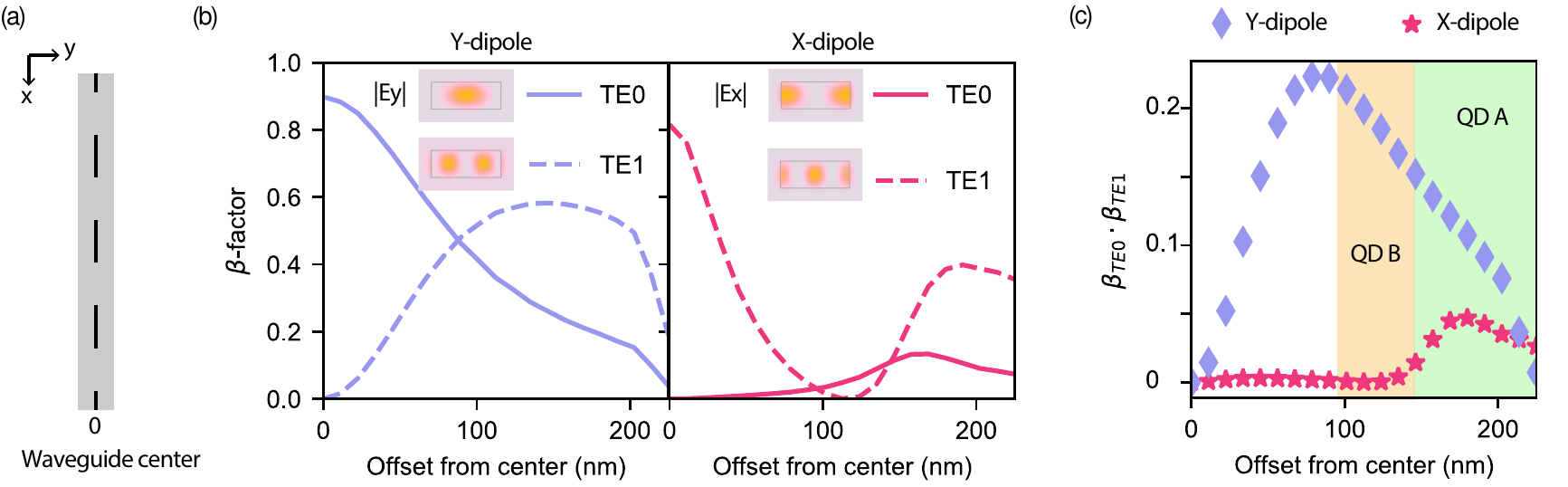}
		\caption{Calculation of the $\beta$ factor for the excitation mode TE1 and collection mode TE0 from FDTD simulation. (a) For the simulation, a dipole source is swept across a $450$ nm-wide waveguide and can be polarized along the $x$-axis or $y$-axis. (b) Calculated $\beta$-factor for the $y$-dipole (left) and  $x$-dipole (right). Relevant mode profiles are shown as insets. (c) The product of the calculated $\beta$-factors of the TE1 and TE0 indicates the probability of exciting and collecting each dipole. Areas which match the dipole intensity ratio of the QDs measured in the main text are highlighted.}
		\label{fig:beta}
	\end{figure*}

	In the main text, the two resonant QDs shown in Fig.~{1}(c) present a second dipole. For QD$_A$, the second dipole is pronounced and is approximately half the first dipole in intensity. For QD$_B$, the second dipole is strongly suppressed. The different behavior in excitation and coupling  of the emission to the waveguide modes can be explained by the 
	$\beta$-factor, which is the ratio between the power coupled to a given mode and the total power radiated by the dipole. Following the method in Ref.~\cite{Rotenberg:17}, we calculate the power radiated in the TE0 and TE1 waveguide modes by performing a three-dimensional finite-difference time-domain \cite{fdtd} calculation carried out using MEEP open-source software package \cite{meep}. To do so, a dual-mode waveguide of width $450$ nm is defined and the position of a radiating dipole, either oriented along the $x-$axis or $y-$axis, is swept across the waveguide width, as shown in Fig.~\ref{fig:beta}(a). The power coupled into the TE1 and TE0 modes, whose mode profiles are shown as inset for the relevant electric field component, are simulated and give the $\beta-$factor for each dipole, shown in Fig.~\ref{fig:beta}(b). In Fig. \ref{fig:beta}(c), we compare the product of $\beta_\text{TE0}\cdot\beta_\text{TE1}$ for each dipole orientation, which indicates the probability of exciting and collecting each dipole. From the intensity ratio of the two dipoles from the RF data of the main text, we estimate that QD$_B$ is offset by approximately $100-125$ nm from the center of the waveguide, while QD$_A$ is much closer to the edge, i.e. around $175$ nm off-center. {For both quantum dots, no Purcell enhancement is expected, as slow-light effects are negligible in nanobeam waveguides.}
	
	\section{Power series under cw-resonant excitation}

\begin{figure*}[h!]
		\includegraphics{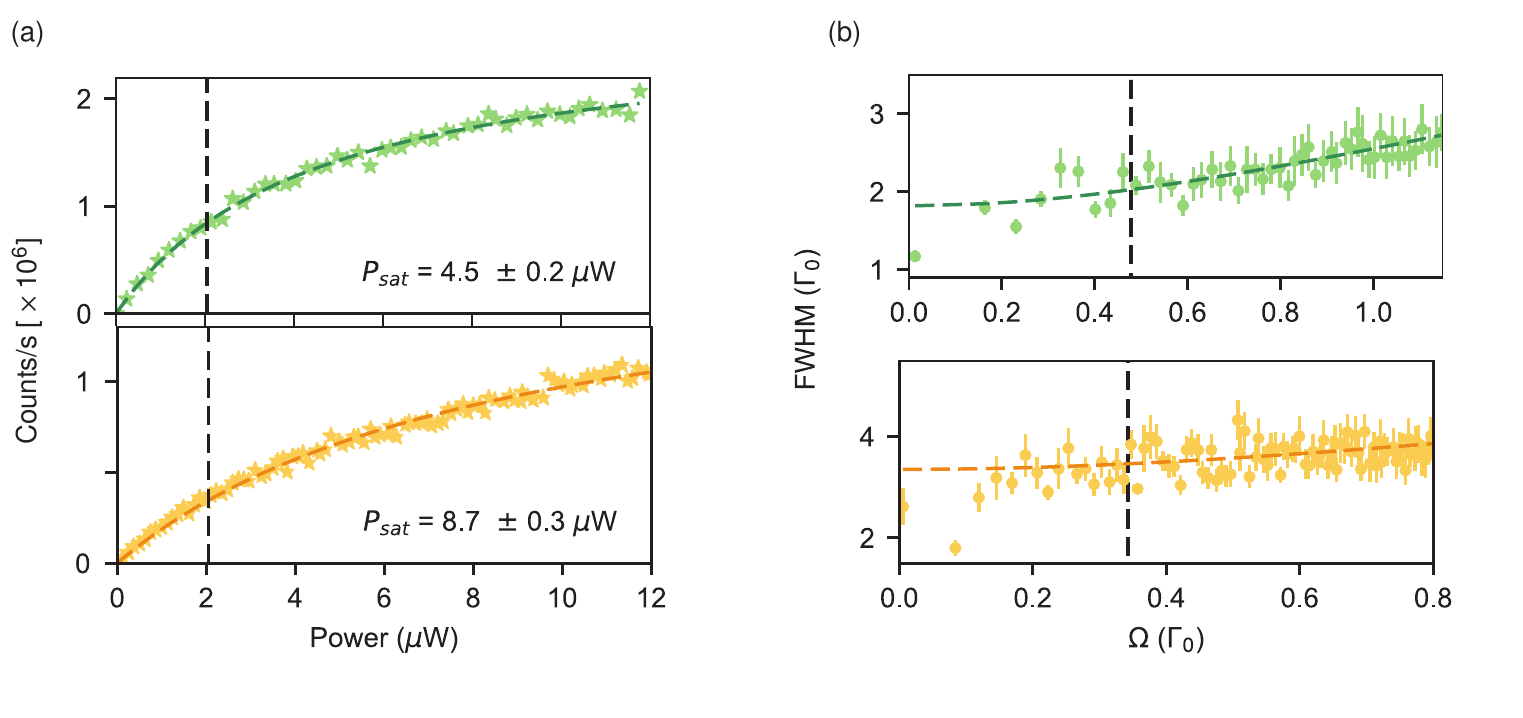}
		\caption{Effect of excitation power on intensity and linewidth. (a) Peak intensity of each QD (QD$_A$, top and QD$_B$, bottom) recorded as a function of power show saturation behavior, fitted with Eq.\ref{eq:saturation}. (b) Lorentzian full-width half-maximum (FWHM) as a function of the Rabi frequency. The black dashed lines indicate the power/Rabi frequency at whihch the experiment in the main text is performed. }
		\label{fig:power}
	\end{figure*}

\begin{figure*}[h!]
		\includegraphics{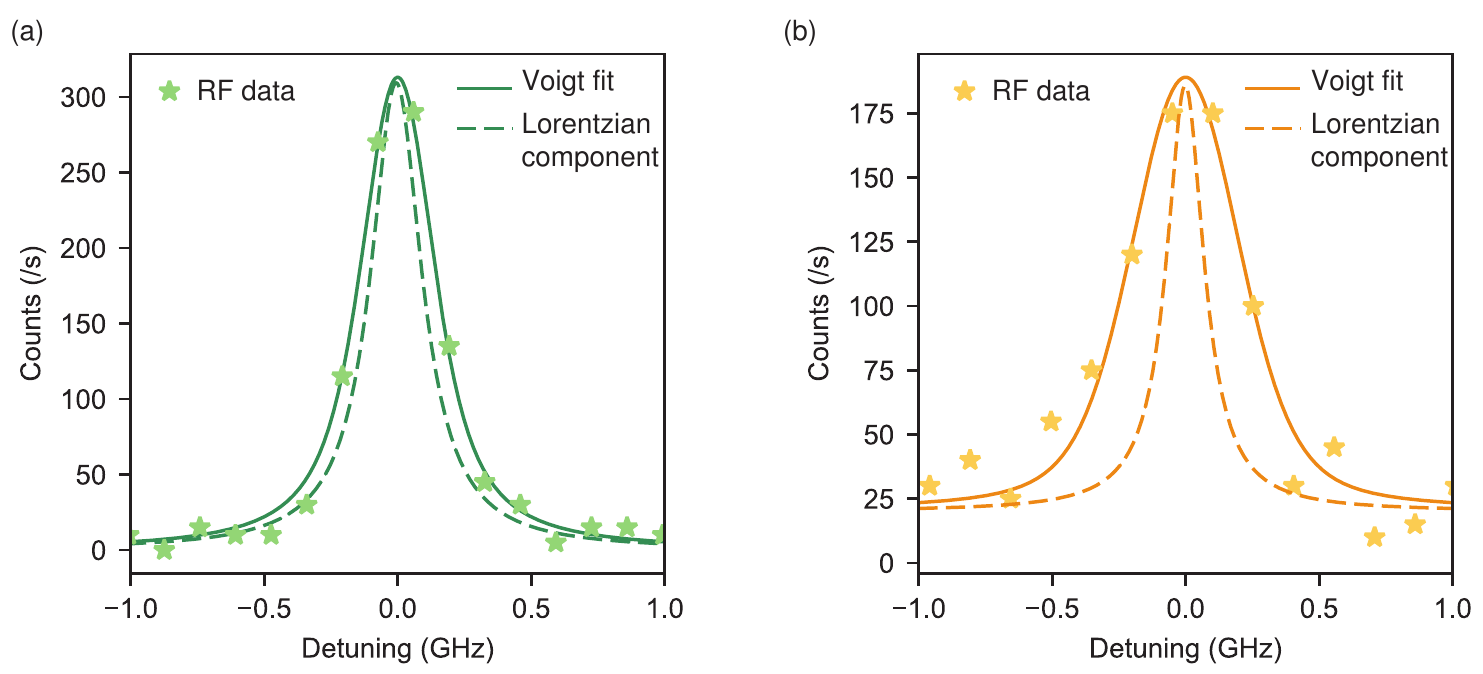}
		\caption{Fit of the low power RF linewidth to a Voigt function. The random noise experienced by QD$_A$ and QD$_B$ follows a distribution with standard deviation $\sigma_A = (68 \pm 4)$ MHz and $\sigma_B = (163 \pm 25)$ MHz, respectively.}
		\label{fig:lowpower}
	\end{figure*}

	In this section, we investigate the effect of the laser power on the intensity of the resonance fluorescence and on the linewidths of the RF spectra. Figure~\ref{fig:power}(a) shows the intensity of QD$_A$ (top, green stars) and QD$_B$ (bottom, orange stars) from the main text saturates as the excitation power increases.
	\newline
	The saturation is fitted to
	\begin{align} 
		\label{eq:saturation}
		I(P) = I_\infty\frac{1}{1+P_{sat}/P},
	\end{align}
	shown as the green (orange) dashed curves, while the power at which the experiment is performed in the main text is shown as the black dashed curve. The power indicated is the value sent towards the sample, without taking into account the transmission of the optics in the cryostat nor the coupling to the PDG and TE1 mode, meaning that the effective saturation power at the QD is in fact significantly lower. Given the radiative decay rate $\Gamma$ of each QD, and by comparing Eq.\ref{eq:saturation} to the population of the excited state of a two-level system driven at resonance \cite{Muller}, the power axis can be translated to the Rabi frequency at which the QD is driven as
	\begin{align} 
		\label{eq:Omega}
		\Omega(P)=\frac{\gamma}{\sqrt{2}}\sqrt{\frac{P}{P_{sat}}}.
	\end{align}
	The Lorentzian full width half-maximum for both QDs is measured as a function of power and is shown in Figure~\ref{fig:power}(b) as a function of the Rabi frequency. This indicates a power broadening, fitted with
	\begin{align} 
		\label{eq:FWHM}
		\Gamma^\text{RF}(\Omega) = \sqrt{\Gamma^2+2\Omega^2}+b,
	\end{align}
	where the constant $b$ accounts for broadening mechanisms beyond power broadening and $\Omega$ is converted to radians/s. 
	This effect is further characterized by fitting the low power RF linewidth to a Voigt function, as shown in Fig.~\ref{fig:lowpower}. The Lorentzian contribution to the linewidth is fixed to the lifetime-limited value (dashed curves) and the standard deviation for the Gaussian distribution of the random noise leading to further linewidth broadening is extracted. 
	
	\section{Laser background}
\begin{figure*}[h!]
		\includegraphics{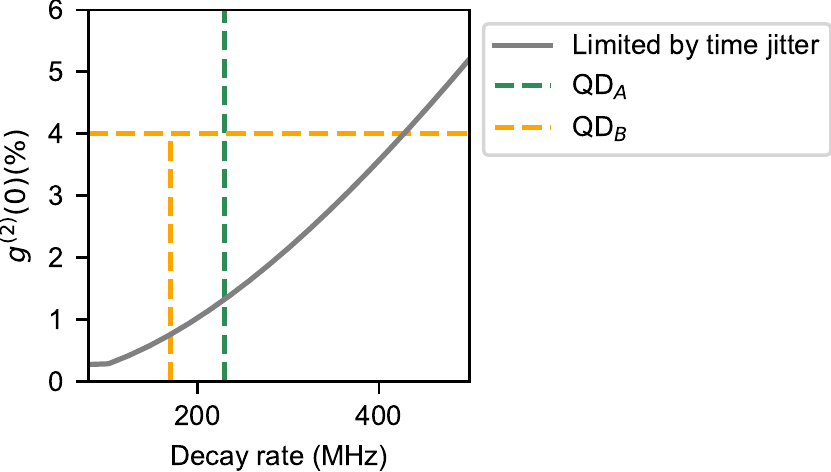}
		\caption{Calculated second-order correlation function evaluated at $\tau=0$ as a function of emitter decay rate, convoluted with the detector response. Given the decay rate of QD$_A$ (green dashed curve) and QD$_B$ (orange dashed curve), the measured $g^{(2)}(0)$ is higher than what it would be if it was only limited by the time jitter of the detectors.}
		\label{fig:g2_IRF}
	\end{figure*}

	In the experiment, the laser background is evaluated by comparing the total RF signal, when the QDs are resonant with the laser, to the detected signal when both gate voltages are set to $1 $ V, meaning that no QD charge states are excited. We obtain an impurity $\xi$, defined as the inverse of the laser suppression, of $\xi_1=0.033$ and $\xi_2=0.017$, for QD$_A$ and QD$_B$ respectively. We can estimate the second-order correlation function as $g^{(2)}(0)=2\xi-\xi^2$\cite{impurity}. The discrepancy between the predictions ($g^{(2)}_A(0)=0.06$ and $g^{(2)}_B(0)=0.03$) and the measured values in the main text can be explained by the fact that the laser background is measured as a single resonance fluorescence scan on both QDs ($40$ s), whereas the second-order correlation function is an average over a longer time period ($10$ min). To address the limitation from the time jitter of our detectors, characterized by a $226$ ps full-width half-maximum (FWHM),  we show in Fig.~\ref{fig:g2_IRF} 
	the expected value of $g^{(2)}(0)$, should it be only limited by the time jitter, calculated for different decay rates with a fixed Rabi frequency of $\Omega = 0.3 \Gamma$. We see that even for the low value of $g^{(2)}(0)$ measured for QD$_B$ (orange dashed curve), the time jitter is not the limiting factor, demonstrating that the laser leakage is the only limitation to the  $g^{(2)}(0)$ value measured in the experiment in the main text.
	
	\section{First-order coherence function and fit parameters of the two-photon quantum interference}
\begin{figure*}[h!]
		\includegraphics{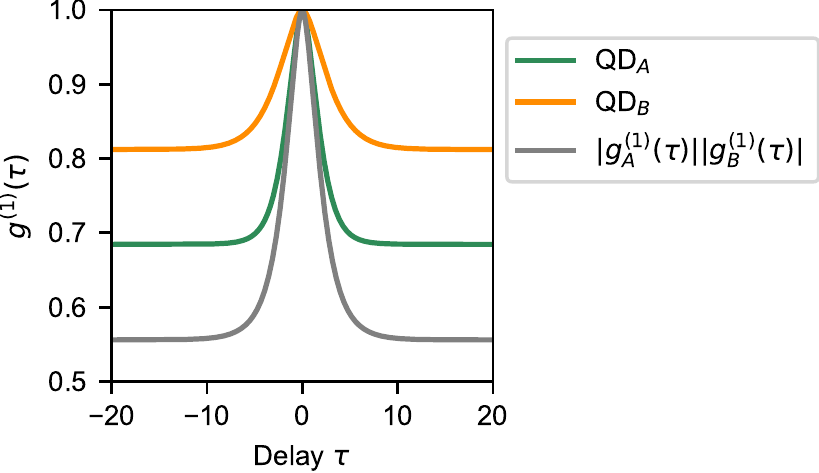}
		\caption{First-order coherence function calculated with $\gamma_{A}/2\pi= 233$ MHz, $\Omega_A=0.48$, $\gamma_{B}/2\pi= 167$ MHz and $\Omega_B=0.34$. The product of the absolute value, involved in the calculation of $g^{(2)}_{\parallel}(\tau)$ in the main text, is shown in gray. }
		\label{fig:g1}
	\end{figure*}

	To predict the HOM visibility, we calculate the first-order correlation function $g^{(1)}(\tau)$  from the resonance fluorescence of a two-level system under continuous-wave excitation, assuming no dephasing, given by \cite{scully_zubairy_1997}
	
	\begin{align} 
		\label{eq:G1}
		G^{(1)}(\tau) = \frac{\Omega^2}{\gamma^2+2\Omega^2}\left[\frac{\gamma^2}{\gamma^2+2\Omega^2}+\frac{1}{2}e^{\frac{\gamma \tau}{2}}+e^{\frac{-3  \gamma \tau}{4}}\left(\cos{(\mu \tau)}\frac{1}{2}\frac{2\Omega^2-\gamma^2}{2\Omega^2+\gamma^2}-\sin{(\mu \tau)}\frac{1}{4\mu}\frac{-5\gamma\Omega^2+\gamma^{3}/2}{2\Omega^2+\gamma^2}\right)\right],
	\end{align}
	with
	\begin{align} 
		\label{eq:mu}
		\mu = \sqrt{\Omega^2-\frac{\gamma^2}{16}}.
	\end{align}
	The normalized first-order coherence function, given by $g^{(1)}(\tau)=G^{(1)}(\tau)/G^{(1)}(0)$, is plotted in Fig.~\ref{fig:g1}, with $\gamma=\gamma_{A}$ ($\gamma_{B}$) from the main text and $\Omega_A \ (\Omega_B) = 0.48\gamma_{A}\  (0.34\gamma_{B})$, from Fig. \ref{fig:power} for QD$_A$ (QD$_B$). 
	\newline
	The calculated first-order coherence is then included in the model of $g^{(2)}_{\parallel}(\tau)$ presented in the main text in Eq.~(2), however due to the excitation scheme (resonant excitation with medium Rabi frequency) we note that $\lim_{|\tau|\rightarrow\infty} g^{(1)}(\tau)  \neq 0 $, in opposition to previous works, where non-resonant excitation is used \cite{Sandoghdar} or high Rabi frequencies in RF \cite{Gerber_2009}. We therefore need to introduce a normalization constant $R$, as shown in Eq.~(2) in the main text, left as a free parameter for the fit of $g^{(2)}_{\parallel}(\tau)$, yielding $R = {2.02}$. This constant is then fitted to $R ={1}$ when performing the ensemble average of the model with $\sigma_{\Delta\omega} = 177$ MHz.
	\newline
	
	\section{ Previous works with two quantum dots}
	{In this section, we provide a summary of significant previous works demonstrating operation of two quantum dots and compare them against our current demonstration of independent operation of two waveguide-integrated quantum emitters. We have gathered the main achievements of these works in Table \ref{t1}, where we put an emphasis on the integration into waveguide. The novelty of our work clearly stands out from previous demonstration of two-photon quantum interference between different quantum emitters.}
	\begin{table}[!h]
		\centering
		\begin{adjustbox}{max width=\textwidth}
			
			\renewcommand{\arraystretch}{2.8}
			
			\begin{tabular}{|c|c|c|c|c|c|c|c|}
				
				\hline
				Ref. & \shortstack{Waveguide \\ integration} &  \shortstack{Emission \\ wavelength} & $g^{(2)}(0)$ $^a$ &  \shortstack{TPQI \\ peak visibility } & \shortstack{Photon \\ indistinguishability }  & \shortstack{Excitation \\ method }   & \shortstack{Tuning \\ mechanism }  \\ [0.5ex]\hline\hline
				\textbf{This work} & \textbf{Yes}&\textbf{$\mathbf{942}$ nm}& \textbf{$\mathbf{0.04\pm0.02}$}&$\mathbf{(79\pm2)\%}$&\textbf{N.A}.&\textbf{CW, resonant} &\textbf{Stark effect}\\ \hline
				\shortstack{You et al. (2022)\\ \cite{lu300km}} &No&\shortstack{\ \\$893$ nm + \\ frequency-conversion\\ to $1583$ nm} &$0.051(1)$&N.A.&$(67\pm2)\%$ (raw)&Pulsed RF & \shortstack{Frequency\\conversion} \\ \hline
				\shortstack{Zhai et al. (2022)\\ \cite{Liang2QD}} &No&$750$-$800$ nm&$0.01\pm 0.001$&N.A.&$(93\pm0.8)\%$ (raw)&Pulsed, resonant&Stark effect \\ \hline
				\shortstack{Ellis et al. (2018)\\ \cite{Ellis}} &\shortstack{No\\ (butt coupling)}&$936$ nm&$0.23$&$54\%$&N.A.&CW, p-shell&Stark effect \\ \hline
				\shortstack{Weber et al. (2018) \\  \cite{Weber2019}} &No&\shortstack{ \ \\$904$ nm + \\frequency conversion\\ to $1550$ nm}&N.A.&$(29\pm3)\%$&N.A.&Pulsed, resonant &\shortstack{Frequency\\conversion} \\ \hline
				\shortstack{Reindl et al. (2017)\\ \cite{Reindl}} &No&$780$ nm&$0.016\pm0.006$&$(51\pm5)\%$&$(71\pm9)\%$&\shortstack{Pulsed,\\phonon-assisted}&Strain tuning \\ \hline
				\shortstack{Patel et al. (2010)\\ \cite{Patel2010}} &No&$943$ nm&$0.05$&$(33\pm1)\%$&N.A.&CW, p-shell&Stark effect \\ \hline
			\end{tabular}
			
		\end{adjustbox}

		\caption{Summary of previous studies with two remote quantum dots.}
		\label{t1}
	\end{table}
	\newline
	\footnotesize{$^a$ Only the lowest value of $g^{(2)}(0)$ is reported.}